# Efficient Broadband Terahertz Generation from Organic Crystal BNA Using Near Infrared Pump

Hang Zhao[1,2], Yong Tan[1,2], Tong Wu[1,2], Gunther Steinfeld[3], Yan Zhang[1], Cunlin Zhang[1], Liangliang Zhang[1], Mostafa Shalaby[*1]

1. Beijing Adv. Innovation Center for Imaging Tech. and Key Lab. of Terahertz Optoelectronics, CNU, Beijing 100048, China
2. Beijing Key Lab. for Precision Optoelectronic Measurement Instrument and Tech., BIT, Beijing, China
3. Swiss Terahertz Research-Zurich, Technopark, 8005 Zurich, and Park Innovaare 5234 Villigen, Switzerland

We are reporting on terahertz generation from organic crystal BNA using a 1.15 to 1.55 μm near infrared pump. We observed a very large emission spectrum extending up to 7 THz, compared to 2.5 THz from Ti:Sa pump in previous reports. The maximum optical-to-THz conversion efficiency in our experiment was 0.8% at 1 kHz repetition rate and without saturation leading to a peak electric field of 1 GV/m. Our results show pronounced phase matching privilege for intense terahertz generation using a pump in the 1.15 to 1.25 μm range where high energy near infrared pump sources operate.

Intense terahertz (THz) technology have seen rapid progress in the past decade. The main driver for that is the fast expansion of applications ranging from a fundamental understanding of matter (such as ultrafast structural dynamics in THz pump- X ray probe spectroscopy [1-3]) to applications (such as electron acceleration[4]). Intense terahertz can be generated from different sources such as large scale electron accelerators [5] and laser techniques. However, the laser-based approach is the most commonly used because of its flexibility and accessibility [6-8].

From lasers, intense THz can be generated from both laser plasma [9] and crystals [6]. It's noteworthy to mention that nearly all recent breakthroughs in intense THz source technology have been realized by engineering the generation techniques [10] or through a better understanding of existing materials' properties using well-known laser technologies and materials [6]. Laser plasma is an emerging field with extensive research, but intense plasma-based THz sources have rarely been used in nonlinear terahertz spectroscopy. The present hurdle in such direction is the low conversion efficiency. In that context, organic crystals rose to prominence because of their high efficiency and broad bandwidth (when compared to the standard inorganic Lithium Niobate crystal platform [10]). The spectrum of the THz pulse generated using Lithium Niobate crystal is mainly restricted to the sub-1 THz range, but in many experiments, such as ultrafast structural dynamics, broad bandwidth (2-10 THz) is highly demanded [11, 12]. In that spectral range, DAST crystal [13] (and its derivative DSTMS [6]) is the benchmark where record THz peak field of 83 MV/cm has been shown [6]. Such materials are very efficient for generation with a 1.55 μm pump from Optical Parametric Amplifiers (OPA) with optical to THz conversion efficiency in the 2-3% range [6].

However, there are still technological limitations. First, the nonlinear absorption leads to rapid saturation of conversion efficiency and higher absorption in the THz spectral range. Moreover, although THz generation phase matching is spectrally broad, extending down to 1 μm range, it becomes narrower, below 1.3 μm. This is unfortunate because present high energy near infrared pump lasers operate below 1.3 μm, such as Cr.Frosterite (1.25 μm) [14] and ytterbium lasers (1.05 μm) [15].

In the search for new materials, organic crystal BNA (N-benzyl-2-methyl-4-nitroaniline) has recently attracted attention as a THz emitter [16]. BNA is an old material, invented in 1997 [17-21], but its application for intense THz generation has only been recently demonstrated using conventional Ti:Sa 800 nm lasers. The output THz pulse shows a spectral bandwidth of up to 2.5 THz [16], spectral peak around 1 THz, and a conversion efficiency of 0.25% . In this work, we present BNA as a candidate for efficient THz generation from near infrared (NIR) laser systems.

Our experimental setup is based on a 1 kHz optical parametric amplifier (OPA) pumped by 3.5 mJ, 800 nm, 35 fs, Ti:Sa laser system. The output of the OPA NIR signal is tunable in the range of 1.15 to 1.6 μm. The BNA crystals are pumped by the OPA beam and the THz is collinearly generated. Residual NIR is filtered out using three low pass filters with out-of-band rejection ratio better than 0.1%. The generated THz beam is expanded and then focused using a set of three off-axis mirrors (Fig. 1a). The detection was done using either a Golay cell (MTI Inc.), electro-optical sampling (100 μm GaP) or RIGI (www.swissterahertz.com) uncooled micro-bolometer camera. We used six BNA crystals (www.swissterahertz.com) with variable thicknesses of: 200 μm; 700 μm; 1.1 mm; 2 mm; 2.3 mm; 2.9 mm.

---

[1] Corresponding author: shalaby@swissterahertz.com

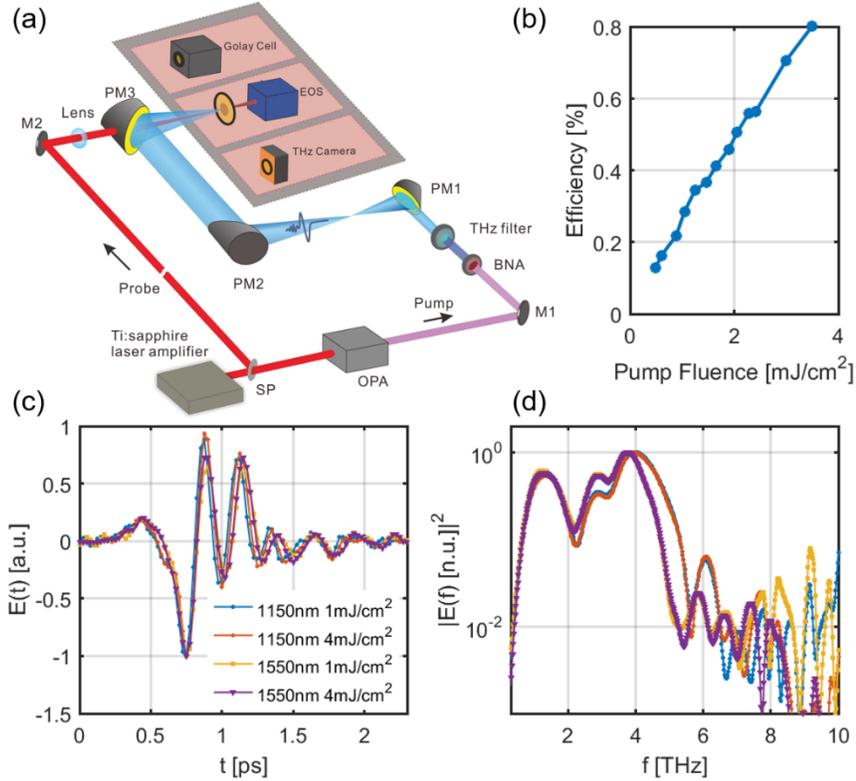

**Figure 1**: (a) Experimental setup of THz generation measurements. (b) Optical to THz conversion efficiency for BNA crystal with a thickness of 1.1 mm. (c) Temporal profiles and (d) the corresponding spectra generated for pump wavelengths of 1150 nm and 1550 nm and for two different pump fluences.

Figure 1c & 1d show the temporal profile and emitted spectrum from the 1.1 mm-thick crystal at a pump fluence of 1 mJ/cm$^2$ and of 4 mJ/cm$^2$ and pump wavelengths of 1150 nm and 1550 nm. The temporal profile features a chirped few cycle THz pulse. The detected spectrum extends from 0.1 to 8 THz, limited by the GaP crystal detector bandwidth. However, the main spectral contents are in the 1-5 THz range. This is much wider than the spectrum obtained with Ti:Sa 800 nm pump which previously reported to extend up to 3 THz with spectral peak around 1 THz [16]. This finding is a direct reflection of the phase matching conditions of BNA [18]. We corrected all the spectra presented in the paper using the GaP detector response function [22]. We did not observe big change in either the temporal profile or spectrum between the two pump fluences. This implies negligible nonlinear absorption even with the maximum fluence we used in our experiment. In terms of conversion efficiency, we show in Fig. 1b the conversion efficiency for different pump fluences. We did not observe saturation in the conversion efficiency up to the maximum fluence used (4 mJ/cm$^2$). The conversion efficiency was strictly linear with the pump fluence, as predicted from a typical optical rectification with minimal nonlinear absorption (thus consistent with the measurements in Fig. 1c and 1d). We attribute that to minimal nonlinear absorption due to the high transparency of BNA in that optical regime compared with conventional DAST/DSTMS systems as well as other yellow crystals (TMS) [23-24]. We did not observe damage or degradation of the crystals performance over a few weeks of continuous testing and storage at ambient room temperature conditions. The damage threshold was found to be better than 10 mJ/cm$^2$ at the laser repetition rate (1kHz with and optical chopper, that is 500 Hz) and central pump wavelength of 1200 nm. We studied the dependence of the generated THz radiation on both the crystal thickness and pump wavelength. In order to gain deeper insight on the phase matching properties, we show the output THz temporal profile for various crystals thicknesses for a pump wavelength of 1150 nm (Fig. 2a), 1250 nm (Fig. 2b), and 1550 nm (Fig. 2c). For 200 μm-thick crystals at 1150 nm (Fig. 2a), the temporal profile shows nearly 1.5 cycle pulse oscillating at an approximate period of 220 fs (4.5 THz). As the thickness increases toward 2.9 mm, such oscillations become more pronounced (several cycles), and superimposed on low frequency (1 THz) contents which become stronger toward thicker crystals. This directly corresponds to two phase matching regimes around 1 and 4 THz [18]. This is directly manifested in the corresponding spectra shown in Fig. 2d, for a pump wavelength of 1150 nm.

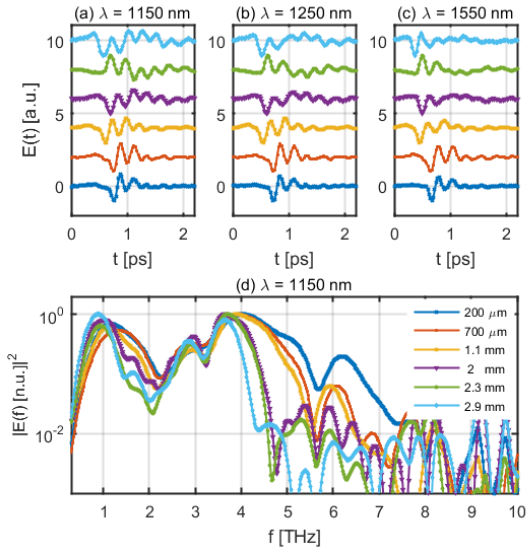

**Figure 2**: Temporal profiles of the generated THz from different crystals at a pump wavelength of (a) 1150, (b) 1250 nm, and (c) 1550 nm. (d) The generated THz spectrum for a pump wavelength of 1150 nm and different crystal thicknesses.

As the pump wavelength increases toward 1250 nm (Fig. 2b) and 1550 nm (Fig. 2c), the low frequency (sub 2 THz) contents are hardly affected. However, the higher frequency contents consistently decrease. Such phase matching properties are the opposite of what common organic crystals (DAST/DSTMS/OH1) should show, where phase matching gets worse and generation bandwidth gets narrower as the pump wavelength decreases from conventional telecommunication/ OPA wavelength of 1550 nm to shorter wavelengths. For example, the main spectral density from DSTMS, shift from 4 THz [6] to 2 THz (1250 nm) [14] using the same crystal. This gives BNA a big application advantage toward upscaling (a present technological challenge) where up to 35 mJ pulse energies can be reached from Cr.Frosterite lasers at 1250 nm [13] compared to fewer mJ from OPA systems available at 1550 nm [6]. Moreover, the phase matching and our experimental results hint at even better generation properties in terms of spectral width and conversion efficiency towards the 1050 nm pump, where very large pump energies are becoming more accessible [15]. We performed calculations (supplementary materials) of the generation phase matching and obtained good agreement with the experimental measurements.

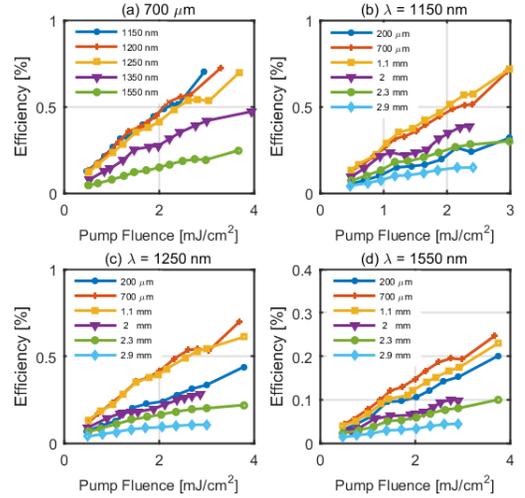

**Figure 3**: (a) Dependence of the THz generation efficiency on the pump wavelength in the 700 μm-thick crystal. Thickness dependence of the generation efficiency is shown for different pump wavelengths of (b) 1150 nm, (c) 1250 nm, and (d) 1550 nm.

The qualitative phase-matching picture depicted in the temporal generation properties of Fig. 2 can be quantified by energy measurements. Fig. 3a shows the generation efficiency of the 700 μm-thick crystal for different pump wavelengths and fluences. There is marginal change in the efficiencies between 1150 nm, 1200 nm and 1250 nm. However, the efficiency monotonically drops when the pump wavelength increases toward 1350 nm and 1550 nm pump wavelengths. This is consistent with the conclusions from Fig. 2. We chose a relatively thin crystal for this comparison to minimize the influence of the thickness. To understand the effect of the thickness, we show the thickness-dependent efficiency for pump wavelengths of 1150 nm, 1250 nm, and 1550 nm in Fig. 3b, 3c, and 3d. The three sets of measurements converge toward the same conclusion that the most efficient thickness of generation is in the 700 μm to 2 mm range. Thinner crystals are less efficient (efficiency is a linear function of the crystal length within the effective generation length). Thicker crystals are less efficient due to the phase mismatch. It is noteworthy to mention that we refer here to the total amount of energy within the 18 THz low pass filter bandwidth. Phase matching is strongly frequency-dependent. Fig. 2 shows that the 4 THz oscillations become stronger as the crystal thickness increases (relative to the background overall THz signal). So, in applications where the spectral contents around 4 THz is of interest, thicker crystals may be more suitable. Optimum thickness would depend on the desired central frequency and spectral bandwidth.

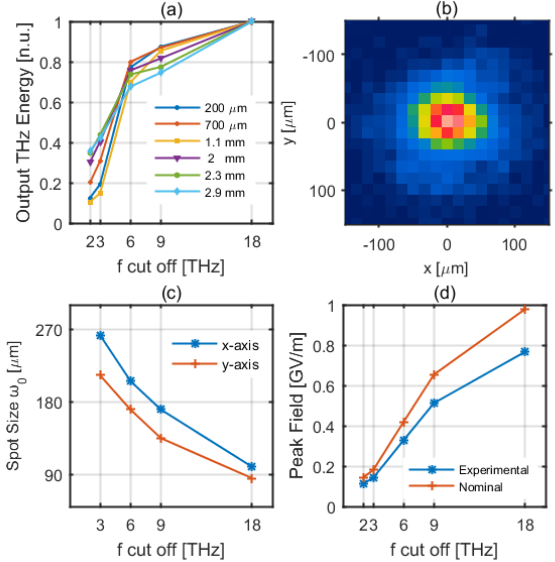

**Figure 4**: (a) Dependence of the generated THz spectral contents on the crystal thickness for a pump wavelength of 1250 nm. (b) Spatial profile of the focused THz beam using 18 THz LPF. It was by RIGI microbolometer camera (www.swissterahertz.com). (c) THz spot size along the horizonal (x) and vertical (y) axes. (d) Peak THz electric field as measured using Kerr effect in diamond (experimental) and after correction for the losses on the THz LPF (nominal). (b), (c), and (d) are measured with a pump wavelength of 1300 nm.

In order to get better understanding of the variation of spectral contents with pump wavelength and thickness, we performed parametric study as shown in Fig. 4. We measured the generation efficiency (at a fluence of 4 mJ/cm$^2$) of the different crystals for pump wavelengths of 1250 nm (Fig. 4a), 1150 nm, 1250 nm, and 1550 nm (supplementary materials). We used a set of low pass filters with cut-off frequencies at 2 THz, 3 THz, 6 THz, 9 THz, and 18 THz. In the case of 1150 nm, 1200 nm, 1250 nm, the main spectral contents are found to be in the 3-6 THz range. As the thickness increases, the contribution of the sub-3 THz to the overall efficiency monotonically increases. The contribution of the spectrum above 6 THz to the overall spectrum varied from 12% (1150 nm pump, 200 µm) to 33% (1250 nm, 2.9 mm). In contrast, the pump with 1550 nm leads to a strong variation in the spectral contents in regard to crystal thickness. This confirms the above discussion of inferior phase matching conditions at such long wavelengths (compared with a shorter wavelength pump). Finally, we show the focused THz image (taken with a RIGI micro-bolometer camera) using the 700 µm-thick crystal and a pump wavelength of 1300 nm (Fig. 4b). We acquired these images using different filter cut-offs. The extracted spot size radii are shown in Fig. 4c. (because the sensitivity of micro-bolometer sensors generally increases with the increase of frequency, and high frequencies correspond to smaller focused spot sizes). Finally, we estimated the peak electric field by measuring the optical Kerr effect in diamond following ref. [13, 25]. We used 300 µm-thick diamond and obtained a maximum phase shift of 32 mrad at a pump wavelength of 1300 nm. This corresponds 0.77 GV/m. This number is corrected for the losses on the THz low pass filters to 1 GV/m nominal peak field.

In conclusion, we have shown that BNA is a potentially suitable THz emitter for intense source upscaling using high energy NIR lasers. The obtained main spectral contents extend up to 7 THz and the conversion efficiency is 0.8 %. We did not observe saturation of conversion efficiency or the effect of nonlinear absorption at the maximum fluence used in our experiment. The damage threshold was found to be more than 10 mJ/cm$^2$ at 1200 nm pump wavelength. The nominal peak electric field was 1 GV/m.

**Supplementary Material**

See supplementary material for more details on the phase matching calculations.

**Acknowledgement**

This work was supported by Beijing Natural Science Foundation (Grant No. JQ18015).